\documentclass[aps,preprint,amsmath,amssymb]{revtex4}

\usepackage{amsmath,amssymb}
\usepackage{bm}
\usepackage{graphicx}
\newcommand{\be}{\begin{eqnarray}}
\newcommand{\ee}{\end{eqnarray}}

\begin{document}

\title{Magneto-Chiral Kerr effect with Application to the Cuprates}

\author{Vivek Aji, Yan He$^1$ and C.M. Varma}

\affiliation{Department of Physics, University of California, Riverside, CA}
\date{\today}
\begin{abstract}
Rotation of polarization of light on transmission and reflection at materials with time-reversal breaking (Faraday and Kerr effects, respectively) have been studied for over a hundred years. We add to such phenomena by studying optical properties of {\it magneto-chiral} states which are states with loop currents such that there is finite Hall effect at zero applied magnetic field. Qualitatively new features arise in reflection and transmission in such a state. This state is shown to be induced in underdoped cuprates given their observed {\it magneto-electric} loop-current order and certain lattice symmetries. These results  explain the observation of Kerr effects with unusual properties and help further confirm the nature of the symmetry breaking in underdoped cuprates.
\end{abstract}
\maketitle

\section{Introduction}

A very interesting experimental development in the novel physics in underdoped cuprates is the discovery of an {\it unusual} Kerr effect \cite{kerr-kap} in underdoped cuprates. The unusual part of the observation is that for a given sample and experimental conditions, the sign of the rotation of the polarization angle is the same on opposite surfaces of the sample, while in the usual Kerr effect the rotation angle must reverse. We will call this the {\it Kapitulnik-Kerr (KK)} effect. In BISCCO-2201 \cite{kerr-bsco} and HgBa$_2$CuO$_{4+\delta}$  \cite{kerr-hgco}, the temperature of onset of the effect $T_{KK} (x)$ is within the experimental uncertainty consistent with $T^*(x)$ deduced from other measurements, including time-reversal breaking observed by neutron scattering for the latter \cite{neutrons}.
The sensitivity of the experiment gives clear evidence of a phase transition even though the detected Kerr rotation corresponds to an effective magnetic moment which is less than $10^{-5} \mu_B$/unit-cell. The smallness of this effect compared to the order deduced by the polarized neutron experiment $(O(10^{-1}) \mu_B$/unit-cell) suggests that the time-reversal breaking observed in KK is an effect induced by the principal order parameter detected by neutrons. 
In several underdoped samples of $YBa_2Cu_3O_{6 + \delta}$  \cite{kerr-ybco},  $T_{KK} (x)$ is consistently lower than $T^*(x)$ but the two head towards zero at the same $x$, the quantum-critical point. A Kerr effect \cite{kerr-lbco} and (the symmetry equivalent) zero-field Nernst effect \cite{nernst-lbco} are also observed in La$_{2-x}$Ba$_x$CuO$_4$ below a specific temperature but  deduction of $T^*$ in it is uncertain.

The unusual phenomena in underdoped cuprates is a central aspect of the mystery of the high temperature superconducting cuprates. Increasing experimental evidence has been adduced which is consistent with the suggestion \cite{cmv-prb97} that there is a transition to an unusual state at $T^*(x)$, below which all thermodynamic and transport properties change. The specific state suggested is a {\it Magneto-electric} (ME) state which breaks
 time-reversal  through orbital current loops within each unit-cell in a pattern that breaks inversion but preserves translational symmetry. Signatures consistent with such a state have been found in four different families of cuprates by polarized neutron scattering \cite{neutrons}. In one of them the suggested dichroic Angle resolved photoemission (ARPES) experiments were also  consistent with such an order. Recent ultrasound experiments \cite{ultrasound} as well as magnetization measurements \cite{leridon} give clear evidence of a thermodynamic phase transition at $T^*(x)$. But as explained below, such a loop-current order itself cannot have a Kerr effect. 
 
Usually Kerr effect is due to ferromagnetic order.  But sensitive magnetization measurements \cite{leridon} rule out onset of such moments to values  greater than about $10^{-7} \mu_B$/unit-cell. An ingenious proposal \cite{orenstein} using variants of the symmetry of the magneto-electric order parameter gives a Kerr effect without ferromagnetic order, but the specific symmetry elements required are not consistent with more recent neutron scattering results \cite{NS}. It also does not give the unusual newly discovered aspect of the KK effect. We present here the physical ideas and calculations leading to the Kerr effect with the observed unusual feature which is also not ferromagnetic and not in conflict with any equilibrium data that we are aware of. 

A Kerr rotation is equivalent to having a finite antisymmetric imaginary part of the dielectric tensor or equivalently off-diagonal or Hall conductivity $\sigma_{xy}$. Four loop-current states, in different point-group symmetry are possible \cite{ASV} in the two-dimensional three orbital/unit-cell cuprate model without breaking translational symmetry. As pointed out by Fradkin and Sun \cite{sun-fradkin} and further elaborated \cite{he-moore-cmv}, one of these possible loop-current states is a  state with a finite Anomalous Hall effect, i.e. $\sigma_{xy} \ne 0$ in zero magnetic field \cite{haldane}.  Such a state breaks time-reversal but breaks chirality rather than inversion as in the ME state observed by neutrons. It preserves the product of time-reversal and chirality. We shall call such a state a {\it Magneto-chiral state} (MC) and show that it has the observed KK effect. We also show that for lattice distortions or potentials of a specific point group (or lower) symmetry, the ME state consistent with observations in neutron scattering must necessarily induce the MC state. We suggest that the coincidence of  $T_{KK}$  and $T^*$ is due to the fact that BISCCO-2201 and HgBa$_2$CuO$_{4+\delta}$ have lattice distortions already present at $T^*(x)$ but that in $YBa_2Cu_3O_{6 + \delta}$ , they set in at a temperature below $T^*(x)$. This is consistent with recent experiments which show structural distortion at below about $T_{KK}(x)$. 

This paper is organized as follows: We begin by briefly describing the ME and the MC states. This is followed by the proof that in presence of specified lattice distortion, the MC state must accompany the ME state. We then calculate $\sigma_{xy}$ for this state to give an estimate for it in relation to experiments. Finally we show by an extension of the methods used in Ref.(\onlinecite{LLP}) for Kerr effect due to time-reversal symmetry breaking and the Faraday effect due to time-reversal preserving gyrotropy that the MC state combines features of both such that it has the special new features of Kerr effect observed by Kapitulnik et al. 

\section{The Magneto-Electric State and the Magneto-chiral State}

\begin{figure}
\includegraphics{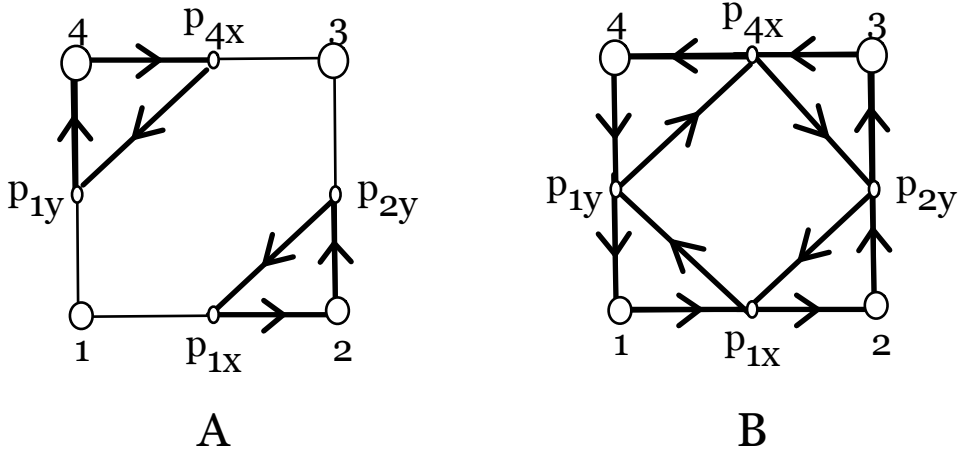}
\caption{The possible magneto-electric Loop current (A) and the Magneto-chiral current (B) Orders in underdoped cuprates}
\label{fluxes}
\end{figure}

The {\it magneto-electric} loop current state in cuprates is described by the order parameter  ${\bf \Omega}_{{\hat{\bf x}'}}$ in each unit-cell, where
\be 
{\bf \Omega}_{{\hat{\bf x}'}} = \int_{unit-cell} d^2 r \Big({\bf L}({\bf r}) \times {\bf r}\Big)
\ee
${\bf L}({\bf r})$ is the orbital magnetic moment at a point ${\bf r}$ measured from the center of each cell.
The current loops in a unit-cell  leading to the two orbital-magnetic moments in each unit-cell  are shown in Fig. (\ref{fluxes}A). Such a state breaks time-reversal and inversion, and preserves only one of the four reflection symmetries of the square lattice, that in the direction $\hat{\bf x}' = \frac{1}{\sqrt{2}}(\hat{\bf x}+\hat{\bf y})$.  (There also exists an equivalent possible state ${\bf \Omega}_{{\hat{\bf y}'}};  {\hat{\bf y}'}\cdot {\hat{\bf x}'} =0.$) As discussed already \cite{sun-fradkin, he-moore-cmv}, such an order parameter has $\sigma_{xy} =0$. Although each ${\bf k}$ state carries a Hall-current, states ${\bf k}$ and $-{\bf k}$ states carry Hall current in opposite directions because the {magneto-electric state} breaks inversion,while preserving the product of inversion and time-reversal. The net effect is zero anomalous Hall current.

As shown in (\onlinecite{ASV}), the three orbital model for cuprates in fact gives five translational symmetry preserving loop current states in distinct point group symmetries for the square lattice. The Loop ordered states were deduced in a mean-field approximation by first  expressing the nearest neighbor interactions $H_{nn}$ in the Cu-O and the O-O bonds in terms of the biquadratics of the five possible closed loops of currents in a cell. As derived in Eq. (D10) of  Ref. (\onlinecite{ASV}), the gauge-invariant, spin-singlet, and translation-preserving parts of the interactions
\be
\label{ints}
H_{nn} &= &\sum_{<R,R'>}  V_{pd} n_{d,R}( n_{x, R'} + n_{y, R'}) + V_{pp} n_{x, R}n_{y, R'} 
\ee
can be written in terms of current loops:
\be
\label{curr-loops}
\sum_{i} \Big(-\frac{V_{pd}}{16}\Big)\Big[|{\bf \Omega}_{i, s}|^2 + |{\bf \Omega}_{i, x'}|^2 + |{\bf \Omega}_{i, y'}|^2 +\frac{1}{2}|{\bf \Omega}_{i, (x^2-y^2)}|^2\Big] - \Big(\frac{V_{pp}}{8}\Big) |{\bf \Omega}_{i, \bar{s}}|^2 , 
\ee
In (\ref{ints}), $<R,R'>$ denote the O nearest neighbors of a given Cu site as well as the O nearest neighbors of a given O site. In (\ref{curr-loops}), $i$ labels the unit-cells and the labels $s, \bar{s}, x', y'$ and $x^2-y^2$ denote loop currents in different point group representations.  All the five current loops ${\bf \Omega}_{i, \alpha}$ \cite{footnote2} are depicted in Fig. (5) of Ref. (\onlinecite{ASV}); we have reproduced the current loop state ${\bf \Omega}_{i, x'}$ and ${\bf \Omega}_{i, \bar{s}}$ in Fig. (\ref{fluxes}). As is clear from this figure, the latter  breaks time-reversal and is chiral. We refer to such a state as {\it Magneto-chiral} (MC). Such a
 a state has a finite Hall effect in zero magnetic field (AHE) \cite{sun-fradkin, he-moore-cmv,palee}.
The effective one-particle Hamiltonian for such a state is $H(AHE) = \sum_{\bf k} H_{\bf k}(AHE)$ is
\be
\label{AHE}
H_{\bf k}(AHE)=\left(\begin{array}{ccc}
0 & its_x & its_y\\
-its_x & 0 & t's_xs_y+irc_xc_y\\
-its_y & t's_xs_y-irc_xc_y & 0
\end{array}\right)
\ee
with $s_{x,y} = \sin(k_{x,y}a)/2; c_{x,y} = \cos(k_{x,y}a)/2$.  In Fig. (\ref{fluxes}B), the flux through the central square is $\propto r/t$ and that in the four surrounding triangles is $\propto -\frac{1}{4} r/t$.

\section{Coupling between Magneto-electric and Magneto-chiral States}

We now show that ME state must be accompanied by the MC state in a lattice with a specific symmetry. The required symmetry of the lattice is such that it has the same broken inversion and reflection symmetries which are broken by ${\bf \Omega}_{i, x'}$ in the "perfect" square lattice. Specify such broken lattice symmetries by ${\bf \epsilon}_{\hat{x}'}$. Then it follows that 
 an invariant in the free-energy density of the form
\be
\alpha {\bf \epsilon}_{\hat{x}'}\cdot {\bf \Omega}_{i, x'}{\bf \Omega}_{i, \bar{s}}
\ee
is allowed. Therefore a state with $\langle{\bf \Omega}_{\bar{s}}\rangle
 \ne 0$ is mandated if $\langle{\bf \Omega}_{\hat{x}'}\rangle \ne 0$. Its magnitude is given by $\alpha \chi_{AHE} {\bf \epsilon}_{\hat{x}'}\cdot\langle{\bf \Omega}_{\hat{x}'}\rangle$,  where the (positive) free-energy of the AHE  in the "perfect" lattice is $(1/2 \chi_{AHE})|{\bf \Omega}_{\bar{s}}|^2$.

\begin{figure}
\includegraphics[width=0.6\textwidth]{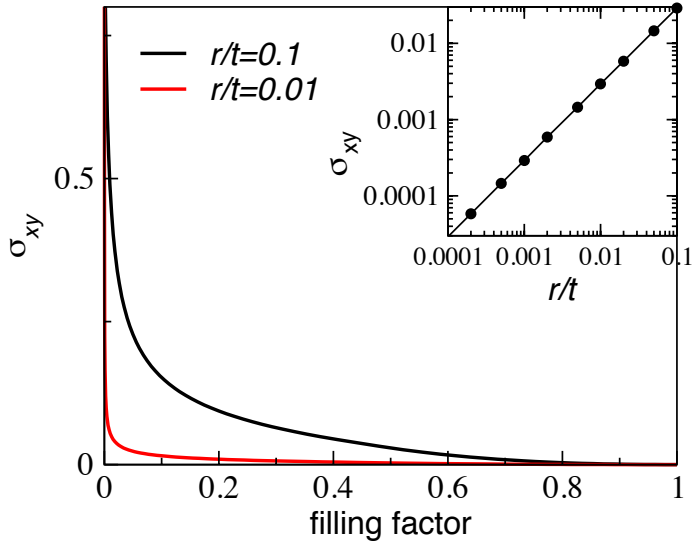}
\caption{Berry curvature as a function of filling for the conduction (and with a change of sign) of the valence band for indicated values of the effective flux coefficient $r/t$. The inset gives the Anomalous Hall coefficient deduced, $\sigma_{xy}/(e^2/h)$ for a conduction band filling of near $1/2$, representative of the underdoped cuprates.}
\label{Chern}
\end{figure}

We can show how  $\langle{\bf \Omega}_{\bar{s}}\rangle
$ is induced by $\langle{\bf \Omega}_{\hat{x}'}\rangle $ by generalization to a distorted lattice of the procedure by which (\ref{curr-loops}) was obtained.
Consider a Cu-O lattice such that the effective inter-site potentials (and the transfer integrals) in the Cu-O bonds and/or the O-O bonds have the point-group symmetry of a given domain of the magneto-electric state. This could happen with just the movement of the two O atoms within a unit-cell without the Cu changing  their square lattice configuration or more complicated arrangements could be envisaged \cite{footnote}.
 The simplest realization of the idea is if, for example, the top-right and the bottom-left  Cu-O and O-O bonds have different  potentials $V_{pd} \pm \delta V^{x'}_{pd}, V_{pp} \pm \delta V^{x'}_{pp}$. It is obvious that with such perturbations the five current loops of Eq. (\ref{curr-loops}) are not mutually orthogonal.  Specifically, one generates to leading order the terms,
 \be
 \label{new-terms}
\frac{\delta V^{x'}_{pd}}{32} \Big({\bf \Omega}_{\bar{s}} {\bf \Omega}_{\hat{x}'}+ {\bf \Omega}_{\hat{x}^2-\hat{y}^2} {\bf \Omega}_{\hat{y}'}\Big).
 \ee
This means that if ground state already has $\langle {\bf \Omega}_{\hat{x}'}\rangle \ne 0$,  a finite $\langle{\bf \Omega}_{\bar{s}}\rangle$ must be generated due to the symmetry broken by $\delta V^{x'}_{pd}$, so that there is an anomalous Hall effect. If the ground state already has 
$\langle {\bf \Omega}_{\hat{y}'}\rangle \ne 0$, a finite $\langle  {\bf \Omega}_{\hat{x}^2 - \hat{y}^2}\rangle$ is generated and there is no anomalous Hall effect. The situation changes correspondingly for a perturbation $\delta V^{y'}_{pd}$. 

\section{Calculation of Anomalous Hall Conductivity}

The zero-field $\sigma_{xy}$ for the Magneto-chiral state given by (\ref{AHE}) is now calculated. For $t'/t<1$, the Chern number of the 3 bands are $-1$, $0$ and $1$ from the bottom respectively. The Chern number is not sensitive to $r/t$ for any full band.
For partially filled conduction bands, we can integrate the Berry curvature up to the chemical potential. The resulting Chern number as a function of band-filling is shown in Fig. (\ref{Chern}) for $t'/t=0.1$, $r/t=0.1$ and $r/t = 0.01$. The point to note is
 that the Chern density in the (non-overlapping) valence and the conduction band are concentrated more and more sharply towards the top of the former and the bottom of the latter as $r/t$ decreases. Therefore for a given filling of the conduction band the observable $\sigma_{xy}$ (and its consequences), which sums over the contributions of the bands integrated up to the chemical potential \cite{haldane} decrease rapidly towards $0$ as $r/t$ decreases. 
 
 In the inset of the figure we show the deduced $\sigma_{xy}/(e^2/h)$ for a conduction band filling of near $1/2$, representative of the underdoped cuprates. The result that $\sigma_{xy} \propto r/t$ can be also be easily derived analytically, for a chemical potential not close to the edge of the bands, by expanding the wave-function in $r/t$ and calculating the Chern integral and noting that the correction is of $O(r/t)^3$. This fails near the edge of the band as for a full band the chern-number is independent of $r/t$.
 
 \section{Special Features of the Kerr Effect}
 
We now discuss the new features brought about in a Kerr effect experiment by the MC state because a finite $\sigma_{xy}$ alone does not fully characterize its chirality breaking aspect.  The relation between the applied electric field $E_i$ and the displacement $D_j$ for an applied magnetic field or a time-reversal breaking order given by (See, for example, Eq. (101.6) of Ref. (\onlinecite{LLP})) by:
 \be
 \label{LLP101.6}
 E_i = \eta_{ik}' D_k + i({\bf G} \times \langle {\bf D}\rangle)_{i}
 \ee
 where the dielectric function $\epsilon_{ik} = \epsilon_{ik}' + i \epsilon_{ik}''$, ${\bf \eta} \equiv {\bf \epsilon}^{-1}$ and ${\bf G}$ is related to the dielectric tensor and  to the order parameter. This is allowed because ${\bf G}$ is an axial time-reversal odd vector so that $i({\bf G} \times {\bf D})$ is a polar time-reversal even vector just as ${\bf E}$. For the simple situation of an external magnetic field or a spatially homogeneous time-reversal breaking order parameter $g_k$, 
 \be
\label{G}
 G_i = -\frac{1}{|Det ~ \bf \epsilon|} \epsilon_{ik}' g_k ,
 \ee
 $i  g_k$ specifies the magnitude and the direction of time-reversal odd vector. $i g_k $ in the text-book examples is due to a magnetic field or magnetization uniform over the sample. The expectation value put on ${\bf D}$ is to indicate that the effect is due to polarization produced due to the change in wave-functions which is linear in the applied  electromagnetic field and which is finite in any unit-cell. This hardly needs stating in the usual case but is crucial for the effects we consider below.
 
 Let us generalize this result to time-reversal breaking states that also break chirality, particularly of the kind  shown in Fig. (\ref{fluxes}B) for Cuprates in which the magnetization consist of a part which is pointing up in one part of the unit-cell and down in the other and is zero averaged over the unit-cell. 
 These magnetizations are produced by a pair of {\it inequivalent} current loops in each unit-cell. This is more obvious if one draws the unit-cell as rectangular in shape than as shown in Fig. (\ref{Chern}B). One loop is bounded by four O-ions and encloses a Cu-ion at its center, the other also is bounded by four O-ions and has no ion inside it. Each loop is described mathematically by its chirality and the direction of the effective magnetic field through it \cite{ASV}:
 \be
 \label{g}
i {\bf g}_{\mu}(x,y) = ig_{\mu}(x \partial/ \partial y - y \partial/ \partial x)_\mu\hat{{\bf z}}.
 \ee
 Here $ig_{\mu} \hat{z}$ signifies the magnitude and the direction of the field through the loops $\mu=1,2$ and the operator $(x \partial/ \partial y - y \partial/ \partial x)_\mu$ specifies its chirality. (This operator should be considered to be acting with respect to the center of each loop).  Both the direction of the field and the chirality are reversed in $\mu=1$ compared to $\mu=2$. On applying an external electromagnetic field, the crystal linearly responds by generating a polarization and therefore a contribution to ${\bf D}(x,y)$ in each unit-cell which has the same symmetry as the order parameter. There is then a spatial variation of  ${\bf D}(x,y)$ in each unit-cell of the form $xy(x^2-y^2)$ in one loop, and $-xy(x^2-y^2)$ in the other. This must happen because the chiral operator $(x \partial/ \partial y - y \partial/ \partial x)$ has the same symmetry as $xy(x^2-y^2)$. Since  the two loops are {\it chemically} inequivalent, the the magnitude $g_1 \ne g_2$. So in the presence of an external electro-magnetic field, local field effects induce a pair of chiral variation in  ${\bf D}(x,y)$  inside each unit-cell, the same in each unit-cell for long-wavelength response. It follows that,
 below the loop order transition, 
  \be
 \label{gav}
 g\hat{\bf z} \equiv \sum_{\mu} g_{\mu}\langle(x \partial/ \partial y - y \partial/ \partial x) {\bf D}(x,y)\rangle_{\mu} \hat{\bf z} \ne 0.
 \ee
  when the expectation values are calculated with wave-functions functions changed linearly in the applied electromagnetic field $E_i$. The sum over the two loops in each unit-cell is finite. 
    
The MC loops of Fig. (\ref{fluxes}) occur in two varieties of domains, the one shown and one with the direction of currents reversed.
We note the important fact that the two domains have identical values of $ g\hat{\bf z}$ because reversal of chirality reverses the direction of magnetic field. The special nature of the magneto-chiral effect may be better understood by comparing the above to the case of a chiral polar molecule lying in each (two-dimensional) unit-cell in which case there is no $\hat{z}$ to think about or a ferromagnet when there is no chirality to think about. In either of those cases different domains would have opposite signs of the relevant $g\hat{\bf z}$.

 One can now proceed with the standard methods of calculating electromagnetic propagation, for example, as given in Ref. (\onlinecite{LLP} -Sec.101). Let the refractive index for normal incidence on the Cu-O planes in say the $+{\hat z}$ direction and assuming a tetragonal symmetry be $n_0$ in the state without the broken time-reversal symmetry. \footnote{The orthorhombic symmetry of some of the cuprate crystals only produces corrections to the considerations in this paper, while it is of-course the dominant effect when considering questions of bireferingence}. The refractive indices in the presence of MC order for electromagnetic propagation for anti-clockwise (right-handed) and clockwise (left-handed) circular polarization, looking along the wave-vector,
 are given by
 \be
 \label{ref index}
 n_{\mp} = n_0 \mp g,
 \ee 
 with corresponding polarizations given by 
 \be
 \label{polar}
 D_x = \mp iD_y.
 \ee
 There is then a rotation of linear polarization. The direction of rotation is determined by the sign of $g$. Consider propagation in the $-\hat{z}$ direction. Then without the chirality operator, as for the usual Faraday or Kerr effect, the direction of rotation (with respect to the $- {\hat z}$ axis) is opposite to the  direction of rotation in the first experiment. But the action of the chirality operator operator is now opposite to that for incidence in the $+ {\hat z}$ direction. Therefore the direction of rotation is the same as before. The unusual feature of the Kapitulnik-Kerr effect is thus obtained. One of the consequences of this is that in a Kerr effect experiment, where light travels the same route in two different directions, there would be no effect but for optical absorption proportional to the length. This is just as in purely gyrotropic materials.
 
 Suppose the sample has multiple domains of the ordered phase in the region of incidence. Normally, this would lead to zero effect in either pure Kerr or pure Gyrotropic propagation. In our case, a reversed domain reverses both the chirality and the flux of the order parameter and so the same direction of rotation is to be expected irrespective of the domains because the sign of $g$ is independent of the domain. This is an important point. In any situation of broken symmetry, there are bound to be domains whose characteristic length is much smaller than the spot size (O(1 micron)) of the optical beam. To get a finite effect, it is essential that the sign of the effect be independent of the domain. This also explains another remarkable feature of the experiments - the sign of the effect does not depend on the history of the sample studied \cite{kapitulnik-pc}.

The $\sigma_{xy}$ calculated in Fig. ({\ref{Chern}) used in the usual expression for the rotation angle gives a rough measure measure of the Kerr rotation angle \cite{kerr-kap} because of the loss necessary to get a net effect in propagation in opposite directions. The measured KK rotation angle is of $O(10^{-6})$ radians, compared to $O(1)$ radian for, say Fe. It is hard to estimate the precise number to compare with experiments since many of the parameters, especially the lattice distortions, are not quantitatively known. We may however get such a number from Fig.(\ref{Chern}) using that the principal order parameter measured by neutron scattering is of $O(0.1) \mu_B$/triangle and if the distortion produces $\delta V_{pd}/V_{pd}$ of $O(10^{-2})$.

This discussion of optical properties of magneto-chiral systems may well have more general applications than to the cuprates alone. Note in particular that all the Anomalous Hall states of the Haldane kind \cite{haldane} are magneto-chiral, although to our knowledge this appears to be the first case of realization of such a state. Also any magneto-electric material with inversion symmetry breaking independently may be expected to have unusual optical properties related to those described here. 

\section{Comparison with Alternate Ideas}

Several ideas have been put forth to explain the observed Kerr effect in cuprates. We have already commented on
an earlier work by Orenstein (\onlinecite{orenstein}).Three papers suggesting that the effect is caused by intrinsic chirality have since been put forth. Two of them rely on structural chirality \cite{chiral}. The idea is that some form of charge modulation (i.e. stripes) occurs which is helical with helicity axis normal to the planes. We find it unnatural to ignore that at least in two families of compounds an order parameter which is $O(10^4)$ larger occurs with $T^*(x) \approx T_{KK}(x)$ and that although there is disparity between the two in one of the compounds, both extrapolate to $0$ at nearly the same point $x_c$. An alternate idea \cite{yakovenko} is that the magneto-electric order observed in the cuprates actually rotates among its four possible directions in going from one plane to the other periodically and is therefore chiral. This may be tested by neutron scattering experiments but it has the problem that the magnitude of the order observed by neutron scattering if made chiral is orders of magnitude larger than that required for the observed Kerr effect.

\section{Summary}
This paper has (1) shown a new class of polarization phenomena in reflection and transmission of electromagnetic waves in {\it magneto-chiral} materials, (2) shown that with appropriate lattice symmetry breaking loop current states of the magneto-electric variety induce loop current states (magneto-chiral states) which have anomalous Hall effects, $\sigma_{xy} \ne 0$ for $H=0$, (3) found  that for (multiband) metals with chemical potential not too far from half-filling, the magnitude  $\sigma_{xy}/(e^2/h)$ is expected to be very small, (4) shown that given the observation of the loop current state consistent with magneto-electric variety as the major order parameter in underdoped cuprates, the satellite order of the magneto-chiral kind explains the novel observations of Kapitulnik et al. The two novel features in the experiments are that the same direction of rotation of polarization is obtained on reflection from opposite surfaces and that the domains expected in actual samples do not destroy the effect nor does heating and cooling through the onset temperature.

{\it Acknowledgements}: We acknowledge conversations about the experimental results, several of them before publication with Aharon Kapitulnik and Steve Kivelson, as well as for discussions of the possible explanations. Thanks are also due to Peter Armitage, Patrick Lee and Victor Yakovenko for useful communications. This research is partially supported by NSF under grant DMR-1206298.

$^1$ Present Address: James Franck Institute, University of Chicago


\begin{thebibliography}{99}
\bibitem{footnote2}
There is a change in notatation from Ref.(\onlinecite{ASV}) to this paper. What were being referred to as ${\bf L}$ in that reference are being referred to as ${\bf \Omega}$, reserving the name ${\bf L}$ for orbital angular momenta.
\bibitem{footnote}
There may in fact be small changes in the periodicity of the lattice due to some of the perturbations; for example in BISCCO, the BiO layer is incommensurate with respect to the Cu-O(2) layer and so a small incommensurate distortion of the latter is mandated. This leads to the variations in the interaction energies and the kinetic energies taken into account here. But at the leading level, one need not consider the variation of the Cu-O(2) from its ideal structure.
\end{thebibliography}
\end{document}